\documentclass[12pt]{article}

\usepackage{amsmath,amssymb,amsfonts,amsthm,bbm,cancel,wasysym}
\usepackage{epsfig,graphics,graphicx,epstopdf}
\usepackage{rotating}
\usepackage{colordvi,color}
\usepackage{array,booktabs,colortbl,multirow}

\newcommand{\scriptmath}[1]{{\scriptsize{\mbox{$#1$}}}}

\newcommand{\bfmath}[1]{{\mbox{\boldmath{$#1$}}}}

\newcommand{\ctoprule}{\toprule[0.5mm]}
\newcommand{\cbottomrule}{\bottomrule[0.5mm]}
\newcommand{\cmrule}{\midrule[0.25mm]}

\newcommand{\crowcolor}{\rowcolor[rgb]{0.9,0.9,0.9}}
\newcommand{\whitecell}{\cellcolor[rgb]{1,1,1}}

\newcommand{\be}{\begin{equation}}
\newcommand{\ee}{\end{equation}}
\newcommand{\bea}{\begin{eqnarray}}  
\newcommand{\eea}{\end{eqnarray}}

\newcommand{\refeq}[1]{\mbox{Eq.~(\ref{#1})}}

\newcommand{\hc}{\mathrm{h.c.}} 

\newcommand{\SM}{\mathrm{SM}}
\newcommand{\mt}[1]{\mathrm{#1}}

\newcommand{\units}[1]{~\mathrm{#1}}

\newcommand{\Zprime}{{$Z^\prime$}}
\newcommand{\Zp}{{Z^\prime}}
\newcommand{\zp}{{$Z^\prime$}~}
\newcommand{\Zpbf}[1]{\bfmath{Z^\prime_{#1}}}

\begin{document}


\begin{flushright}
UG-FT-285/11 \\
CAFPE-155/11 \\

August 1, 2011
\end{flushright}
\vspace*{5mm}
\begin{center}

\renewcommand{\thefootnote}{\fnsymbol{footnote}}

{\Large {\bf  Impact of extra particles on indirect $\bfmath{Z^\prime}$ limits}} \\
\vspace*{1cm}
{\bf F.\ del Aguila$^a$}\footnote{E-mail: faguila@ugr.es},
{\bf J.\ de Blas$^b$}\footnote{E-mail: jdeblasm@nd.edu} , 
{\bf P.\ Langacker$^c$}\footnote{E-mail: pgl@ias.edu} \\
and
{\bf M.\ P\'erez-Victoria$^a$}\footnote{E-mail: mpv@ugr.es}

\vspace{0.5cm}

$^a$ Departamento de F\'{\i}sica Te\'orica y del Cosmos and CAFPE,\\
Universidad de Granada, E-18071 Granada, Spain \\
~\\
$^b$ Department of Physics, University of Notre Dame, \\
Notre Dame, Indiana 46556, USA \\
 ~\\
$^c$ School of Natural Sciences, Institute for Advanced Study, \\
Princeton, NJ 08540, USA

\end{center}
\vspace{.5cm}

\begin{abstract}
 
\noindent We study the possibility of relaxing the indirect limits on extra neutral vector bosons by their interplay with additional new particles. 
They can be systematically weakened, even below present direct bounds at colliders, by the addition of more vector 
bosons and/or scalars designed for this purpose. Otherwise, they appear to be robust.  

\end{abstract}

\renewcommand{\thefootnote}{\arabic{footnote}}
\setcounter{footnote}{0}


\section{Introduction}

In the coming years, the Large Hadron Collider (LHC) will explore energy scales up to a few TeV. New physics beyond the Standard 
Model (SM) at these scales could be unveiled, mainly in the form of new particles that give rise to resonances in the $s$ channel. 
One appealing possibility is the production of extra neutral vector bosons \Zprime \footnote{Neutral vector bosons can appear as 
components of different irreducible representations of $SU(2)_L\times U(1)_Y$. We will concentrate on $SU(2)_L\times U(1)_Y$ 
(and color) singlets. This is what is commonly called a \zp boson.} (for a recent review, see \cite{Langacker:2008yv}). These particles 
appear in many extensions of the SM, such as grand unified theories (GUT), scenarios with strong electroweak symmetry breaking, 
theories in extra dimensions and little Higgs models. \zp  bosons stand among the best candidates for an early discovery at the LHC. 
If coupled to quarks and leptons, they can be easily observed as dilepton peaks in the Drell-Yan process 
$q\overline{q}\rightarrow Z^\prime \rightarrow \ell^+\ell^-~(\ell=e,\mu)$. On the other hand, the existence of $Z^\prime$ bosons with adequate couplings could explain the possible discrepancies with the SM predictions at large colliders, such as the anomalies recently found at the Tevatron: a $3.4~\sigma$ discrepancy in the top forward-backward asymmetry for large invariant masses \cite{Aaltonen:2011kc} (see \cite{Jung:2009jz}), or the $3.2~\sigma$ excess in the $W+jj$ distribution around $M_{jj}\sim150\units{GeV}$  \cite{Aaltonen:2011mk} (see \cite{Buckley:2011vc}).

However, other experiments have already placed stringent bounds on \zp masses and couplings. They narrow significantly the 
parameter space available for LHC searches, especially in the early stages of the LHC at 7~TeV center of mass energy and low luminosity. These bounds are of two kinds: direct and indirect. Until recently, the best direct limits on \zp bosons came from searches 
at the Tevatron. The same processes and couplings that are being studied at the LHC are relevant in this case. For sizes of couplings as 
in GUT (SM), the Tevatron typically puts a lower bound of $\sim 800$ GeV (1 TeV) on the \zp masses \cite{Abe:1997fd,Erler:2010uy}. In this regard, it is noteworthy that, with only $\sim 40\units{pb}^{-1}$ of luminosity, the first LHC data allows one to derive limits comparable to those from the Tevatron~\cite{Chatrchyan:2011wq,Aad:2011xp}. Moreover, in some cases LHC bounds are already slightly stronger than the Tevatron ones.
On the other hand, there are strong indirect limits from electroweak precision data (EWPD) obtained from measurements at the $Z$ 
pole (LEP and SLC), at low energies (experiments on parity violation and neutrino scattering), and above the $Z$ pole (LEP 2 and Tevatron) 
\cite{delAguila:2010mx} (see also \cite{Durkin:1985ev,Salvioni:2009mt}). 

The usually quoted \zp limits depend on the implicit assumption that there are no other new particles that could change the analysis. 
However, in most models with extra neutral vector bosons, the \zp is accompanied by other particles, such as additional neutral vectors, 
extra charged vector bosons, exotic fermions (sometimes required for anomaly cancellation), scalars (which may get a vacuum expectation value [vev]), etc. In many 
instances, these additional particles do have an impact on the bounds. Direct limits are relaxed if the \zp can decay into other particles 
beyond the SM, thus increasing its width. This happens, for instance, in some \zp supersymmetric models~\cite{Kang:2004bz}. 
Additional new particles contributing to EWPD will also modify the indirect bounds. Typically, the inclusion 
of more particles just makes the limits more stringent; in this case, the limits obtained from the analysis of a \zp alone are valid as conservative 
limits. But it is also possible that their contributions cancel some effects of the \Zprime, in such a way that the limits 
are relaxed. In this paper we explore this possibility. By looking at the systematics of the cancellations and studying particular examples, 
we shall be able to judge the robustness of the standard \zp limits.

The observable  \zp effects involving the SM matter 
fermions  $\psi$ can be parametrized by the \zp physical mass $M_{\Zp}$, width $\Gamma_{\Zp}$,
mixing $s_{Z\Zp}$ with the $Z$, and couplings $g^\psi$ to the  $5\times 3$ SM fermion multiplets.
 For simplicity, we shall assume family universality 
of the \zp couplings (although similar arguments would apply to nonuniversal scenarios),
yielding eight parameters. Particular models or classes of models impose relations on the couplings, thus reducing the number of independent parameters. For example, if there is only one  Higgs doublet
$\phi$ (as in the SM), and there are no additional particles light enough to affect  $\Gamma_{\Zp}$, then
$\Gamma_{\Zp}$ and $s_{Z\Zp}$ can be computed in terms of the coupling to the Higgs, $g^\phi$, and the
other parameters,\footnote{The Higgs mass $M_H$ affects the partial width for $\Zp \rightarrow Z H$, but this effect is small for $M_\Zp \gg M_H$. The effect of $M_H$ on the radiative corrections to EWPD are more important, as discussed below.} for a total of seven. For example, the mixing  is given by $$s_{Z\Zp} \approx \frac{g^\phi \sqrt{g^2+{g^\prime}^2}}{2} \frac{v^2}{M_\Zp^2}.$$
The fermion and Higgs couplings may also be related by additional assumptions, such as that the standard Yukawa couplings are allowed.
Similarly,  in specific GUT models the couplings to fermions are given by products of fixed charges times a coupling constant $g_\Zp$, 
which is determined by the unification condition. The coupling to the scalar doublet or doublets may be fixed as well. In some cases this suffices to determine the mixing, while in others it  
depends on the ratios of unknown vevs. In the latter case, the GUT \zp model has two free parameters: $M_\Zp$ and $s_{Z\Zp}$.

EWPD are sensitive to the ratios $g^F/M_\Zp \equiv G^F$, where $F$ represents any SM fermion or Higgs field. They also depend on the Higgs mass $M_H$ through the oblique radiative corrections. Including it, the model independent 
fits with one \zp and a single Higgs have basically eight free parameters,\footnote{Because the uncertainties for the other SM parameters are small, we fix them 
to their values at the SM minimum.} whereas fits for GUT models with free mixing have three. The leading corrections to $Z$-pole 
observables depend linearly on the products $G^F G^\phi$. When the fermion couplings are fixed, this requires either large $M_\Zp$ or 
small mixing $s_{Z\Zp}$. On the other hand, the corrections to observables off the $Z$ pole depend also on the combinations 
$G^{\psi_1}G^{\psi_2}$, with $\psi_i$ any fermion multiplet. Therefore, these observables constrain the ratios $G^{\psi}$ even for vanishing 
mixing. For fixed fermion couplings, they put lower bounds on the \zp mass. It is important to observe that the Tevatron and LHC searches 
and constraints depend on the fermion couplings and the mass, but not significantly on the mixing. Therefore, a vanishing mixing does not diminish the 
chances of observation at LHC.

The indirect limits on physics beyond the SM are in general rather 
stringent \cite{Barbieri:2000gf,deBlasThesis}, 
in particular, on extra neutral vector bosons as already mentioned. 
This reflects two facts: the SM provides a quite good description of EWPD, 
and the largest deviations from the SM cannot be significantly accounted 
by such (simple) SM extensions. Hence, new particles, and in particular 
$Z^\prime$s, are typically banished to high scales near a TeV. 
In the following we discuss more {\em complex} scenarios, where the 
new physics conspires to have little effect on EWPD, except maybe to 
accommodate a relatively heavy Higgs.\footnote{As emphasized in Ref. \cite{delAguila:2010mx} the main 
corrections to electroweak observables due to a relatively heavy Higgs 
can be canceled out by the addition of new vector bosons. Thus, 
as shown in figure 9 of that reference an extra vector boson triplet 
$\mathcal{W}^1$ or an adapted singlet $\mathcal{B}$ do balance 
the heavy Higgs contribution.}

In particular, we study the 
possibility of lowering the most stringent indirect limits on popular \zp 
models by adding a second \zp or/and extra scalars. In general, 
we can always lower those limits  with specific additions designed 
for that purpose. Otherwise, the EWPD limits are robust. 
In section \ref{section: Evading EWL} we introduce the effective 
dimension-six operators describing the \zp contributions to EWPD 
observables. The effective Lagrangian approach is especially 
well-suited for the comparison of the contributions of 
any new heavy  physics addition because it provides a basis of operators 
parametrizing any weakly-coupled SM extension. 
Then, after discussing the different \zp effects, we present 
 the numerical analysis in section \ref{section: num_results}. 
We update the present indirect limits on popular $Z^\prime$s with masses 
banished above a TeV, and characterize for each case the \zp addition 
which may largely cancel the popular \zp contributions to 
electroweak observables. In all cases we can lower those 
limits below the present Tevatron and LHC limits  with properly chosen 
$Z^\prime$s and scalars. For comparison, we also study the case of the minimal 
models discussed in \cite{Salvioni:2009mt}, where the $Z$-$Z^\prime$ mixing is not a free parameter. The corresponding 
constraints on the mixing can be somewhat relaxed by adding extra neutral and charged 
vector singlets. Our conclusions are collected in section \ref{Conclusions}.


\section{Evading electroweak constraints}
\label{section: Evading EWL}

We want to examine which kind of additional new physics can relax the limits on \zp bosons from EWPD. This is not straightforward, 
because the new particles that can neutralize the corrections of the \zp to certain electroweak observables may simultaneously increase the 
discrepancy in others. A  convenient way of analyzing the collective effect of several different particles is through their
contribution to the gauge-invariant effective Lagrangian that describes arbitrary extensions of the SM at energies below the masses 
of the new particles. This procedure is more efficient than examining each of the many observables.

To analyze current EWPD it is sufficient to include dimension-four and dimension-six operators: 
$\mathcal{L}_\mathrm{eff} \approx \mathcal{L}_\mathrm{SM} + \mathcal{L}_{6}$. Here, $\mathcal{L}_\mathrm{SM}$ is the SM Lagrangian 
and
\be
\mathcal{L}_6 = \frac{1}{\Lambda^2}\sum_i \alpha_i \mathcal{O}_i,
\ee
where $\mathcal{O}_i$ are gauge-invariant dimension-six operators, $\Lambda$ a scale of the order of the mass of the lightest extra 
particle, and $\alpha_i$ dimensionless coefficients. We will use the complete set of operators in~\cite{Buchmuller:1985jz} 
(see also \cite{Grzadkowski:2003tf}). In a given extension of the SM, the coefficients $\alpha_i$ can be written in terms of the couplings 
and ratios of masses of the new particles. 
The operators that contribute to EWPD can be classified into three groups:
\begin{itemize}
\item Oblique operators, which modify the $Z$ and $W$ propagators.
\item Operators made of scalars, gauge vector bosons (or derivatives), and fermion fields (scalar-vector-fermion [SVF] operators). They contribute to the trilinear 
couplings of two fermions and the $Z$ and $W$ bosons.
\item Four-fermion operators. Only the operators with at least two leptons contribute to EWPD. 
\end{itemize}
At the $Z$ pole, only the oblique and SVF operators give significant contributions. The oblique operators also change the SM prediction 
for $M_W$. The high precision of these measurements strongly constrains the values of the oblique and SVF operator coefficients. 
Therefore, in practice only the four-fermion operators can give important contributions to observables off the $Z$ pole.

A sufficient condition for the new physics to be invisible to EWPD is that $\alpha_i$ vanishes (or is small enough) for all the operators 
contributing to electroweak observables. This condition is in practice also necessary, since different combinations of operators appear 
in the different observables. So, our strategy is simply to look for cancellations that make the relevant $\alpha_i$ vanish:
\be
\alpha_i = \alpha_i^{\Zp} + \alpha_i^{\mathrm{extra}} = 0.
\ee

To start with, we need to know which of the operators that contribute to EWPD are induced by the \Zprime, and the values of their 
coefficients $\alpha_i^{\Zp}$ in terms of the \zp parameters. The following operators contributing to EWPD are generated in a 
generic \zp model at tree level (see~\cite{delAguila:2010mx}\footnote{One can use other operator bases to describe the integration of $Z^\prime$s \cite{Cacciapaglia:2006pk}, which may be more 
convenient for other purposes. We choose to use the standard basis \cite{Buchmuller:1985jz,Grzadkowski:2003tf}, which is well adapted 
to perform global fits.}):
\begin{itemize}
\item One oblique operator ${\cal O}_\phi^{(3)}= (\phi^\dagger D_\mu \phi)((D^\mu \phi)^\dagger \phi)$. After electroweak symmetry breaking this operator, induced by the $Z$-$\Zp$ mixing, contributes 
to the $\rho$ parameter. Its coefficient is proportional to the $T$ parameter. No other oblique parameter appears 
in our operator basis at this order.
\item Five SVF operators, ${\cal O}_{\phi \psi}^{(1)}=(\phi^\dagger i D_\mu \phi )(\overline{\psi} \gamma^\mu \psi)$.
\item Nine four-fermion operators, ${\cal O}_{l\psi}^{(1)}=\frac{1}{1+\delta_{\psi l}}(\overline{l_L}\gamma^\mu l_L)(\overline{\psi_L}\gamma_\mu \psi_L)$, 
${\cal O}_{l\psi}=(\overline{l_L}\psi_R)(\overline{\psi_R}l_L)$, 
${\cal O}_{e\psi}=\frac{1}{1+\delta_{\psi e}}(\overline{e_R}\gamma^\mu e_R)(\overline{\psi_R}\gamma_\mu \psi_R)$, and 
${\cal O}_{qe}=(\overline{q_L}e_R)(\overline{e_R}q_L)$.
\end{itemize}
Our notation is as follows: $\psi$ stands for any fermion multiplet; $\psi_{L,R}$ denote the left-handed (LH) doublets and right-handed (RH)
singlets, respectively; $l_L,q_L$ represent the lepton and quark LH doublets; and $e_R,u_R,d_R$ the RH lepton 
and quark singlets. The SVF and four-fermion operators have two and four flavour indices, respectively, which we have not displayed. 
We have obtained the coefficients of these operators in~\cite{delAguila:2010mx}. The explicit expressions are collected in Appendix B of 
that reference (the \zp is called $\mathcal{B}$ there).
All the coefficients induced by a single \zp have the factorized form
\be
\alpha^\Zp_{FF^\prime} \propto g^F g^{F^\prime}.
\ee
Because, in the universal scenario, there are only six different \zp couplings $g^F$, there are many relations between the fifteen operator 
coefficients.

Let us now examine which types of new particles can give the right contributions $\alpha_i^{\mathrm{extra}}$ to offset, at least partially, 
the \zp coefficients. We discuss, in turn, the different types of operators.

\subsection{Oblique operator}
\label{Obliqueop}

The \zp contribution to the coefficient of the oblique operator ${\cal O}_\phi^{(3)}$ is negative definite: 
$\left(\alpha_\phi^{(3)}\right)^\Zp = -2 \left(g^\phi\right)^2 \leq 0$. The contribution to the $\rho$ parameter has opposite sign, so it is positive 
definite. The same would hold, clearly, for any number of heavy \zp bosons. This effect can be compensated for in several ways.

First, we need not resort to additional new physics. The loop effects of a heavy Higgs (with respect to the ones for a light Higgs, which is 
preferred by EWPD within the SM) give a negative correction to $\rho$ and can counterbalance to a large extent the \zp contribution. In fact, 
if the Higgs were found to be heavy, a \zp extension would be clearly favoured over the SM. This mechanism has been analyzed quantitatively 
in~\cite{delAguila:2010mx}. It should be noted that the heavy Higgs also induces universal SVF operators, contributing to the $S$ parameter, 
so the cancellation is not perfect.

Second, a vanishing $\alpha_\phi^{(3)}$ can be achieved if the \zp is accompanied by a singlet vector boson with hypercharge $Y=1$, such as the one 
that appears in left-right models. This field gives rise, upon electroweak symmetry breaking, to an extra charged vector boson. 
Its main effect in EWPD is a negative contribution to $\rho$, proportional to the square of its coupling to the scalar doublet. In fact, this mechanism 
is at work in any model with a \zp and custodial symmetry. On the other hand, its couplings to RH quarks (and to RH leptons if the neutrinos are Dirac) are constrained by 
measurements of $K^0$-$\bar{K^0}$ mixing, $\beta$ decay, $\mu$ decay,
and weak universality \cite{Beall:1981ze,Nakamura:2010zzi}.

Third, the effects can be canceled or eliminated
if several scalars participate in electroweak symmetry breaking.
There are two distinct effects. One is that the $Z$-$\Zp$ mixing is given for large $M_\Zp$ by
\be
s_{Z\Zp} \sim - \frac{\sum_i t_{3i}g^{\phi_i} \sqrt{g^2+{g^\prime}^2}\left|\left<\phi_i\right>\right|^2}{M_\Zp^2} \label{2higgs}
\ee
for an arbitrary set of scalar fields $\phi_i$ with $\left\{t_i,~\! t_{3i}\right\}$ their weak isospin and third component, respectively. This can be reduced or eliminated by cancellations between $\phi_i$ with
opposite signs for  $ t_{3i}g^{\phi_i}$, which are often present, e.g., in GUT models. A second effect is that
the $\rho$ parameter is modified at the dimension-four level:
\be
\rho^{\mathrm{SM}} \to \rho^{\mathrm{multiple~vevs}} =
\frac{\sum_i \left(t_i^2-t_{3i}^2+t_i\right)\left|\left<\phi_i\right>\right|^2}{\sum_i 2t_{3i}^2\left|\left<\phi_i\right>\right|^2}.
\ee
The \zp contribution can be canceled by adjusting  the vevs of scalars with 
$t_i \ge 1$ and the appropriate $t_{3i}$.

Finally, the simplest possibility is making the \zp scalar coupling $g^\phi$, and thus the mixing, small.
As we have mentioned before, this has no consequences for collider searches.

\subsection{SVF operators}
\label{SVFop}

All the coefficients of SVF operators are proportional to the $Z$-\zp mixing, so the 
simplest way to get rid of them is, again, to have a sufficiently small mixing, either by cancellations, as in (\ref{2higgs}), or by  a small $g^\phi$. It is also possible to cancel a nonvanishing 
\zp contribution with additional particles, as we discuss next.

Extra vectorlike leptons and quarks contribute exclusively to SVF operators via their mixing with the SM fermions 
\cite{delAguila:1982fs,Langacker:1988ur}. 
Using the results of \cite{delAguila:2008pw,delAguila:2000rc}, it is easy to check that for any \zp couplings there 
exist combinations of extra leptons and quarks, with adjusted Yukawa couplings, such that the five net SVF 
coefficients vanish. However, in generic cases the combinations are very contrived: several multiplets in different 
representations of the SM gauge group are required. Moreover, to 
avoid flavour-changing neutral currents a replica of the multiplets is necessary. This is similar to the discussion of scalars in section~\ref{Sec_4leptons}.

A cleaner cancellation is achieved by adding other vector bosons that also mix with the $Z$ boson. According to the 
results in~\cite{delAguila:2010mx}, only the extra vector bosons of vanishing hypercharge that are $SU(2)_L$ and 
color singlets can give contributions to the SVF operators generated by the \Zprime. Hence, we have to resort to 
additional \zp bosons. 
The net contribution of $N$ \zp bosons, including the original one, to the SVF coefficients will vanish if the following 
equations are satisfied:
\be
\frac{\alpha_{\phi \psi}^{(1)}}{\Lambda^2}=-\sum_{n=1}^N G_n^\psi G_n^\phi=0,~~( \psi=l,q,e,u,d),\label{DirectTrilinear}
\ee
with $G_n^F=g_n^F/M_n$ and $g_n^F$, $M_n$ the couplings and masses of the $n$th \Zprime. Already with $N=2$, 
for any fixed couplings $g_1^\psi,\, g_1^\phi$  and fixed mass $M_1$, there are nontrivial solutions that make all these 
coefficients vanish: $G_2^\phi= \pm c_{\mt{SVF}} G_1^\phi$, $c_{\mt{SVF}} G_2^\psi=\mp G_1^\psi$, $ \psi=l,q,e,u,d$, with $c_{\mt{SVF}}$ any real number. 
We call such a companion \zp boson a {\em mirror} \Zprime. 

Of course, the additional \zp bosons increase the deviation in the $\rho$ parameter, but this can be taken care of as 
discussed above. Then, all the $Z$-pole observables will be blind to this pair of \zp bosons. All these \zp bosons also 
contribute to four-fermion observables, as we discuss in the next subsections.

\subsection{Indefinite-sign four-fermion operators}
\label{Sec_2leptons_2quarks}

Scalar and vector bosons both contribute to four-fermion operators. To cancel most of the contributions of the 
\Zprime, the better suited fields are again additional neutral vector bosons, due to their chirality structure and quantum numbers.
However, in the universal scenario the contribution of each \zp to the four-fermion operators ${\cal O}_{ll}^{(1)}$ and ${\cal O}_{ee}$ is negative semidefinite, 
so the different contributions go in the same direction and cannot cancel. The same holds for vector 
bosons with other quantum numbers~\cite{delAguila:2010mx}. Then the question is whether cancelling the coefficients 
of these two operators is possible by introducing scalar fields. We will postpone that analysis to the next subsection, and 
concentrate here on the seven four-fermion operators with coefficients of indefinite sign.

Let us consider again a set of $N$ \zp bosons. The cancellations we are looking for are given,  in this sector, 
by nontrivial solutions to the following system of equations:
\bea
\frac{\alpha_{lq}^{(1)}}{\Lambda^2}=-\sum_{n=1}^{N} G_n^lG_n^q =0 \label{lq},\\
\frac{\alpha_{lu}}{\Lambda^2}=~2\sum_{n=1}^{N} G_n^lG_n^u =0 \label{lu},\\
\frac{\alpha_{ld}}{\Lambda^2}=~2\sum_{n=1}^{N}G_n^lG_n^d=0\label{ld},\\
\frac{\alpha_{le}}{\Lambda^2}=~2\sum_{n=1}^{N}G_n^lG_n^e=0\label{le},\\
\frac{\alpha_{eu}}{\Lambda^2}=-\!\sum_{n=1}^{N}G_n^e G_n^u=0\label{eu},\\
\frac{\alpha_{ed}}{\Lambda^2}=-\!\sum_{n=1}^{N}G_n^e G_n^d=0\label{ed},\\
\frac{\alpha_{qe}}{\Lambda^2}=~2\sum_{n=1}^{N} G_n^q G_n^e=0\label{qe}.
\eea

For given $G^\psi_1$ of the initial $Z^\prime$, the unknowns are the ratios $G_n^\psi$, $n\geq 2$.
In the case of two {\Zprime}s, $N=2$, there are seven equations for five unknowns. In fact, Eqs. (\ref{eu})
and (\ref{ed}) are not independent for $N=2$, and will be automatically satisfied if Eqs. 
(\ref{lq})-(\ref{ld}) and (\ref{qe}) are. However, it is easy to see that Eqs.  (\ref{lq})-(\ref{eu}) have no real solution
unless $G_1^l=0$, $G_1^e=0$, or $G_1^q=G_1^u=G_1^d=0$.
Nevertheless, the EWPD limits can be relaxed if the 
nonvanishing coefficient only enters observables that are less precisely measured. 
With $N=3$, all  seven coefficients of indefinite sign can be zero.

On the other hand, if the $Z$-\zp mixings do not vanish, and assuming no other contributions to four-fermion or SVF operators, 
the system of $Z^\prime$s would have to satisfy at the same time Eqs.~(\ref{lq})-(\ref{qe}) and \refeq{DirectTrilinear}. 
This requires at least four $Z^\prime$s, including the initial one. In view of such a proliferation, it is important 
to keep in mind that each of the extra $Z^\prime$s increases the size of the four-fermion operator coefficients of 
definite sign. 

\subsection{Definite-sign four-fermion operators}
\label{Sec_4leptons}

We address now the operators with four LH leptons, ${\cal O}_{ll}^{(1)}$, and with four RH leptons, ${\cal O}_{ee}$.
Assuming diagonal and universal couplings, the contribution of the $Z^\prime$s to their coefficients is
\be
(\alpha_{ll}^{(1)})_{ijkl}=- \sum_{n=1}^N(G_n^l)^2\delta_{ij}\delta_{kl},~~~~(\alpha_{ee})_{ijkl}
=- \frac{1}{2} \sum_{n=1}^N (G_n^e)^2 \left(\delta_{ij}\delta_{kl}+\delta_{il}\delta_{kj}\right),
\label{B4lops}
\ee
which is indeed negative semidefinite.\footnote{For general (nondiagonal and nonuniversal) couplings only the coefficients with $i=k$ and $j=l$ are negative semidefinite.}
We have displayed in this case the flavour indices, as they will be important in the discussion below. These operators 
contribute to the lepton cross sections and asymmetries at LEP~2, and to purely leptonic low-energy observables, 
such as parity violation in M{\o}ller scattering and neutral current neutrino electron scattering.  
As shown in \cite{delAguila:2010mx}, these operators cannot be canceled by other vector bosons. 
However, we will see that scalar fields can do the job. In general, we need two types of scalars.

The coefficient $\alpha_{ee}$ can be canceled by the contribution of a {\em scalar singlet}  $\varphi$ with 
hypercharge $Y=-2$.\footnote{Note that this kind of scalar fields must not get a vev.} This scalar can only have the following renormalizable interaction with the SM fermions:
\be
\Delta {\cal L}=-\lambda_\varphi^{ij} \varphi^\dagger \overline{e_R^{i~c}} e_R^j+\hc .
\label{singlet_lag}
\ee
The Yukawa coupling matrix $\lambda_\varphi$ is symmetric.
From the interaction above, we see that $\varphi$ has lepton number $L=2$. Integrating it out, the operator 
${\cal O}_{ee}$ is generated, with coefficient
\be
(\alpha_{ee})_{ijkl}=\frac{(\lambda^\dagger_\varphi)^{ki} \lambda_\varphi^{jl}}{M_\varphi^2} .
\label{singlet op}
\ee
The coefficient is symmetric under exchange of the first and third and/or second and fourth family indices, 
just as the coefficient generated by the $Z^\prime$s. This is actually a symmetry of the operator itself. Moreover, 
the coefficient has the right sign required to cancel the effect of the $Z^\prime$s to this operator.
The coefficient $\alpha_{ll}^{(1)}$ can, in turn, be canceled by the contribution of a {\em scalar triplet} $\Delta$ 
with hypercharge $Y=1$. These scalars are well known by their r\^ole as messengers in the seesaw mechanism 
of type II~\cite{Konetschny:1977bn}. For our purposes neutrino masses can be neglected and we can assume lepton number conservation. 
Thus, assigning to $\Delta$ lepton number $L=-2$, its only possible renormalizable interaction with SM fermions is
\be
\Delta {\cal L}=-\lambda_\Delta^{ij} \overline{{l_L^i}^c} i\sigma_2 \Delta^a\sigma_a l_L^j+ \hc .
\label{triplet_lag}
\ee
The contribution to the coefficient of the operator with four LH leptons is~\cite{delAguila:2007ap}
\be
(\alpha_{ll}^{(1)})_{ijkl}=2\frac{(\lambda^\dagger_\Delta)^{ki} \lambda_\Delta^{jl}}{M_\Delta^2}.
\label{triplet op}
\ee
Again, this can neutralize the corresponding \zp contribution. 

There is, however, an important difficulty in realizing these cancellations. Because of the different chiral structure of their 
couplings, the flavour indices of the scalars are crossed with respect to the ones of the vectors. For the relevant 
observables, the first two (or last two) indices correspond to the first family, i.e., to electrons. Fixing these two 
indices in the operator coefficient, for the vector contribution one is left with a symmetric matrix, which in the 
diagonal universal case is proportional to the identity. In the scalar contribution, on the other hand, the coefficient 
reduces to a rank-one matrix. So, the cancellation of the contribution of 
any number of universal \zp bosons with only one singlet and/or triplet scalar 
always leaves nonvanishing off-diagonal scalar contributions. 
Removing them from the purely leptonic four-fermion operators 
requires at least three scalars of each type.

The scalars couplings, in general, lead to lepton 
flavour violating processes, for which there are stringent constraints. For example, the couplings of the triplet must obey 
$|\lambda_\Delta^{e\mu(e\tau)}||\lambda_\Delta^{ee}|/M_\Delta^2<1.2\times 10^{-5} ( 1.3\times 10^{-2})\units{TeV}^{-2}$ from 
$\mu^- (\tau^-)\rightarrow e^- e^+ e^-$, and 
$|\lambda_\Delta^{e\mu}||\lambda_\Delta^{e\tau}|/M_\Delta^2<1.7\times 10^{-2}\units{TeV}^{-2}$ from 
$\tau^-\rightarrow e^- e^+ \mu^-$  \cite{Abada:2007ux}. All these restrictions are satisfied if each scalar couples the 
electron to just one lepton family, i.e., $e$, $\mu$, or $\tau$, and they do not mix. So, for a perfect cancellation 
of the operators ${\cal O}_{ee}$ and ${\cal O}_{ll}^{(1)}$ without flavour violation  we would need to introduce three singlet scalars, $\varphi_{e,\mu,\tau}$, with nonvanishing 
Yukawa couplings $\lambda_{\varphi_e}^{ee}$, $\lambda_{\varphi_\mu}^{e\mu}=\lambda_{\varphi_\mu}^{\mu e}$ and 
$\lambda_{\varphi_\tau}^{e\tau}=\lambda_{\varphi_\tau}^{\tau e}$, respectively, and three triplet scalars, $\Delta_{e,\mu,\tau}$, with nonvanishing Yukawa couplings 
satisfying analogous relations.

However, in the triplet case there is an additional problem: the same coupling 
$\lambda_{\Delta}^{e\mu}$ that enter the  LEP 2 $e^+e^-\rightarrow \mu^+\mu^-$ observables also contributes to $\mu$ 
decay, and hence, indirectly, to other observables such as Cabibbo-Kobayashi-Maskawa universality. In the absence of further new physics 
affecting $\mu$ decay~\cite{delAguila:2009vv}, this coupling is strongly constrained. 
This prevents the required cancellation of the $Z^\prime$ effects on LEP 2 $e^+e^-\rightarrow \mu^+\mu^-$ data.


\section{Relaxing electroweak limits on $\bfmath{Z^\prime}$ models}
\label{section: num_results}

So far we have determined what kind of additional new physics is required to cancel the effects of a 
\zp boson in EWPD. We have seen that an optimal cancellation requires introducing many different 
particles with specific couplings. Such a complicated scenario is, of course, highly unnatural. 
However, in most cases it is possible to weaken the limits significantly using only a few extra particles, 
for instance, adding an appropriate second \Zprime, or scalar singlets and triplets, or both. 
In this section we give several examples  with these two types of additions. 
We consider a few popular \zp 
models and show numerically how the limits are relaxed with the help of these extra particles. 
In particular, we have chosen those examples where current limits exclude the corresponding 
\zp masses below 1 TeV. This includes the $Z^\prime_\chi$, $Z^\prime_I$ and 
$Z^\prime_S$ models, inspired in $E_6$ GUTs, the $Z^\prime_{LR}$ from left-right models, 
the $Z^\prime_R$, whose charges are given by the third component of $SU(2)_R$, and the 
$Z^\prime_{B-L}$ model. The explicit fermionic charges, $Q_{Z^\prime}^\psi$, for these models 
can be found in table \ref{table: ZpCharges}, where we have also included the $Z^\prime_\eta$ charges 
for later convenience. All these charges are related to the ratios $G^\psi_{Z^\prime}$ entering 
in the electroweak fit by $Q_{Z^\prime}^\psi \equiv G_{Z^\prime}^\psi M_{Z^\prime}/g_{Z^\prime}$, 
with $g_{Z^\prime}=\sqrt{5/3} g^\prime\approx 0.46$. For a more detailed description of these 
models and their origin see \cite{Langacker:2008yv}. Finally, we also discuss the effects of extra particles 
in the class of minimal \zp models discussed in \cite{Salvioni:2009mt}. For these models, which we
will denote by $Z^\prime_\mt{min}$, the charges are normalized with $g_{Z^\prime}=\sqrt{g^2+g^{\prime~2}}\approx 0.74$ (see Eq.~(\ref{Jmin}) below). In all cases we 
assume family independent gauge couplings.

\subsection{Extra particle additions cancelling $\bfmath{Z^\prime}$ effects}

For the popular \zp examples the electroweak bounds are typically derived assuming free mixing 
\cite{delAguila:2010mx,Durkin:1985ev}. Thus, the limits on their masses are determined by the constraints 
from observables off the $Z$ pole, i.e., by their contributions to four-fermion operators.\footnote{The corrections to $Z$-pole observables are proportional to the mixing of the $Z^\prime$ with the $Z$ 
boson, $s_{Z\Zp}$, and thus can be adjusted to zero.}

Following the discussion in sections \ref{Sec_2leptons_2quarks} and \ref{Sec_4leptons} such contributions 
can be canceled by adding extra $Z^\prime$s and extra scalar fields. Thus, the addition of such particles 
should allow for weaker bounds in the selected examples. For each of the above mentioned $Z^\prime$s 
we will consider a custom counterpart, denoted by $\overline{Z^\prime}$, with charges  chosen 
to attain (at least partially) such cancellations. The charges for these $\overline{Z^\prime}$ models are also 
shown in table \ref{table: ZpCharges}, below the ones for the corresponding $Z^\prime$ example. Let us 
explain how we choose them in general. First, since we are interested in the contributions to observables 
off the $Z$ pole we can choose the effective Higgs charge of these particles to be zero. For two $Z^\prime$s, 
we can exactly solve six of the equations (\ref{lq}) to (\ref{qe}). Choosing $G_1^{q,u,d}=\mp c_{\mt{4F}} G_2^{q,u,d}$ and 
$c_{\mt{4F}}G_1^{l,e}=\pm G_2^{l,e}$, with $c_{\mt{4F}}$ a real number, we can solve all but (\ref{le}). It is convenient to 
choose $c_{\mt{4F}}$ to be small, which typically ensures that the second \zp would not have been observed in the dilepton searches at the Tevatron and LHC.
A small  $c_{\mt{4F}}$ also minimizes the extra contributions to all four-lepton operators, and, in particular,  avoids 
increasing too much the size of the operators with four LH or four RH leptons, which are definite negative. 
On the other hand, we cannot choose $c_{\mt{4F}}$ to be  too small. Large hadronic ratios would imply either large 
charges that could spoil perturbativity or too low masses, which would go against the effective Lagrangian expansion 
we use (and present limits on extra $Z^\prime$s and other new particles). 
In practice, we increase the hadronic charges by a factor smaller than the one used to suppress 
the leptonic ones, and we restrict to order-one charges.

\begin{table}[t]
\begin{center}
{\footnotesize
\begin{tabular}{c c c c c c c }
\ctoprule
\!\bf{Model}\!& &$l_L$&$q_L$&$e_R$&$u_R$&$d_R$\\
\cmidrule{1-1}\cmidrule{3-7}
&&&&&&\\[-0.4cm]
\crowcolor$\Zpbf{\chi}$&\whitecell &$\frac{3}{2\sqrt{10}}$&$-\frac{1}{2\sqrt{10}}$&$\frac{1}{2\sqrt{10}}$&$\frac{1}{2\sqrt{10}}$&$-\frac{3}{2\sqrt{10}}$\\
\crowcolor &\whitecell &&&&&\\[-0.2cm]
\crowcolor$\bfmath{\overline{Z^\prime}_{\chi}}$&\whitecell &$-\frac{1}{10}Q_{Z^\prime_\chi}^l$&$5Q_{Z^\prime_\chi}^q$&$0$&$5Q_{Z^\prime_\chi}^u$&$5Q_{Z^\prime_\chi}^d$\\
&&&&&&\\[-0.4cm]
$\Zpbf{I}$& &$-\frac{1}{2}$&$0$&$0$&$0$&$\frac{1}{2}$\\
&&&&&&\\[-0.2cm]
$\bfmath{\overline{Z^\prime}_{I}}$& &$-\frac{1}{10}Q_{Z^\prime_I}^l$&$0$&$0$&$0$&$2Q_{Z^\prime_\chi}^l$\\
&&&&&&\\[-0.4cm]
\crowcolor$\Zpbf{S}$&\whitecell &$\frac{2}{\sqrt{15}}$&$-\frac{1}{4\sqrt{15}}$&$\frac{1}{4\sqrt{15}}$&$\frac{1}{4\sqrt{15}}$&$-\frac{2}{\sqrt{15}}$\\
\crowcolor &\whitecell &&&&&\\[-0.2cm]
\crowcolor$\bfmath{\overline{Z^\prime}_{S}}$&\whitecell &$-\frac{1}{10}Q_{Z^\prime_S}^l$&$5Q_{Z^\prime_S}^q$&$0$&$5Q_{Z^\prime_S}^u$&$5Q_{Z^\prime_S}^d$\\
&&&&&&\\[-0.4cm]
$\Zpbf{LR}$& &$\sqrt{\frac 3 5}\frac{1}{2\alpha}$&$\sqrt{\frac 3 5}\frac{1}{6\alpha}$&$\sqrt{\frac 3 5}\left(-\frac{\alpha}{2}+\frac{1}{2\alpha}\right)$&$\sqrt{\frac 3 5}\left(\frac{\alpha}{2}-\frac{1}{6\alpha}\right)$&$-\sqrt{\frac 3 5}\left(\frac{\alpha}{2}+\frac{1}{6\alpha}\right)$\\
&&&&&&\\[-0.2cm]
$\bfmath{\overline{Z^\prime}_{LR}}$& &$-\frac{1}{10}Q_{Z^\prime_{LR}}^l$&$2Q_{Z^\prime_{LR}}^q$&$-\frac{1}{10}Q_{Z^\prime_{LR}}^e$&$2Q_{Z^\prime_{LR}}^u$&$2Q_{Z^\prime_{LR}}^d$\\
&&&&&&\\[-0.4cm]
\crowcolor$\Zpbf{R}$&\whitecell &$0$&$0$&$-\frac 12$&$\frac 12$&$-\frac 12$\\
\crowcolor &\whitecell &&&&&\\[-0.2cm]
\crowcolor$\bfmath{\overline{Z^\prime}_{R}}$&\whitecell &0&0&$-\frac{1}{10}Q_{Z^\prime_{R}}^e$&$2Q_{Z^\prime_{R}}^u$&$2Q_{Z^\prime_{R}}^d$\\
&&&&&&\\[-0.4cm]
$\Zpbf{B-L}$& &$-\frac{1}{2}$&$\frac{1}{6}$&$-\frac{1}{2}$&$\frac{1}{6}$&$\frac{1}{6}$\\
&&&&&&\\[-0.45cm]
\cmidrule{1-1}\cmidrule{3-7}
%
\crowcolor$\Zpbf{\eta}$&\whitecell &$\frac{1}{2\sqrt{15}}$&$-\frac{1}{\sqrt{15}}$&$\frac{1}{\sqrt{15}}$&$\frac{1}{\sqrt{15}}$&$-\frac{1}{2\sqrt{15}}$\\
&&&&&&\\[-0.38cm]
\cmidrule{1-1}\cmidrule{3-7}
%
$\Zpbf{\mt{min}}$& &$-\frac{1}{2}g_{Y}-g_{B-L}$&$\frac{1}{6}g_{Y}+\frac{1}{3}g_{B-L}$&$-g_{Y}-g_{B-L}$&$\frac{2}{3} g_{Y}+\frac{1}{3}g_{B-L}$&$-\frac{1}{3}g_{Y}+\frac{1}{3}g_{B-L}$\\
\cbottomrule
\end{tabular}
\caption{SM fermion charges, $Q_{Z^\prime}^\psi \equiv G_{Z^\prime}^\psi M_{Z^\prime}/g_{Z^\prime}$, with $g_{Z^\prime}=\sqrt{5/3}g^\prime$ ($g_{Z^\prime}=\sqrt{g^2+g^{\prime~2}}$) for the popular (minimal) $Z^\prime$ models discussed in section \ref{section: num_results}. For the 
left-right model $\alpha=1.53$. The charges for the $\overline{Z^\prime}$ models are functions of the corresponding 
$Z^\prime$ charges and chosen for convenience. See the text for details. The charges for the $Z^\prime_\eta$ model 
are also included for completeness.}
\label{table: ZpCharges}}
\end{center}
\end{table}

For the cancellation of the four-lepton operators with definite sign we will use the extra scalars 
$\varphi$ and $\Delta$ introduced in section \ref{Sec_4leptons}. To illustrate the possible effects without introducing
(adjusting) too many extra parameters, we will restrict to a somewhat
minimal scenario. We will consider three scalars singlets,
$\varphi_{\ell}$, each of them coupling
the electron to only one lepton family $\ell = e, \mu, \tau$,
respectively, avoiding
lepton flavour violating constraints in this way.
Moreover, we choose all their couplings in Eq. (\ref{singlet_lag}) to
be equal
$\lambda^{e\ell}_{\varphi_\ell}(=\lambda^{\ell
e}_{\varphi_\ell}) \equiv \lambda_\varphi$,
$\ell=e,\mu,\tau$.
On the other hand, we will only add one $\Delta$ and with a nonzero
coupling only to
electrons, $\lambda_\Delta^{ee} \equiv \lambda_\Delta$,
thus evading $\mu$ decay data constraints.
It must be stressed that the choice of equal $\varphi$
scalar couplings
does not allow for the exact cancellation of the four-lepton operators
resulting from
the $Z^\prime$ integration, as they have a different coefficient relation
by a factor
of two. Thus, $\left(\alpha_{ee}\right)_{eeee}=2\left(\alpha_{ee}\right)_{ee\mu\mu}$ for a $Z^\prime$ with diagonal and universal couplings, whereas $\left(\alpha_{ee}\right)_{eeee}=\left(\alpha_{ee}\right)_{ee\mu\mu}$ for our choice of $\varphi$ couplings (see Eqs. (\ref{B4lops}) and (\ref{singlet op})). If we had chosen the $\varphi$ scalar
couplings to
obtain such a cancellation, the numerical results (limits) below would not
change much
because the fits involve many other data and contributions, with the
exception of
the $Z^\prime_R$ addition as we will comment when discussing this extended model.
Obviously, allowing for many different arbitrary additions and couplings
the $\chi^2$ can be relatively improved, but not its significance.

\subsection{Results and discussion}

We have updated the fit to EWPD for each of the above mentioned popular $Z^\prime$ examples \cite{delAguila:2010mx}, 
and performed new fits to the further extended models. 
For each case we compute the bounds on the $Z^\prime$ mass and the new parameters controlling the 
interactions of the extra particles. The results are presented in table \ref{table: ZpLimits}.\footnote{In general
 we allow for arbitrary $Z$-$Z^\prime$ mixing. In order for this to vanish in the models
considered, however, an extended scalar sector allowing for cancellations between different unknown vevs
is required, as emphasized in section \ref{section: Evading EWL}. In the fits below we take this mixing to be
 a free parameter, except for the minimal models which are discussed at the end, without introducing explicitly the corresponding extra scalars.}
Prior to discussing 
them, let us sketch a few details about the fits. We largely follow \cite{delAguila:2010mx} and refer there 
for further details. There have been several updates in the experimental data considered in that reference. 
We use the updated values for the top mass \cite{:1900yx}, the SM Higgs direct searches limits from Tevatron 
\cite{:2010ar}, and the value for the five-quark contribution to the running of $\alpha_\mt{em}$ 
\cite{Teubner:2010ah,Davier:2010nc}. Other changes include the value of the $W$ width \cite{:2010vi} and 
some updates in the low-energy data \cite{Nakamura:2010zzi}. With these improvements the SM fit gives 
$M_H=125\units{GeV}$~\footnote{Without including the direct limits the best fit value for the Higgs mass still passes 
the barrier of 100 GeV, $M_H=105^{+32}_{-26}\units{GeV}$, getting closer to the LEP 2 
exclusion bound of $114\units{GeV}$. This shift is due to the slightly larger value of the new 
top mass but mainly to the new determination of $\Delta \alpha_\mt{had}^{(5)}\left(M_Z^2\right)$ 
\cite{Davier:2010nc}.}, 
$m_t=173.6\pm1.0\units{GeV}$, $M_Z=91.1876\pm0.0021\units{GeV}$, 
$\alpha_S\left(M_Z^2\right)=0.1184\pm0.0007$, and 
$\Delta \alpha_\mt{had}^{(5)}\left(M_Z^2\right)=0.02751\pm0.00008$. 
At the minimum, $\chi^2_\SM=158.5$ for a total of $207$ degrees of freedom. As explained in the 
introduction, we fix all the SM parameters except the Higgs mass, which is left free, to these best SM fit values. 
Finally, the $95\%$ C.L. limits (two-dimensional regions) presented in this section are obtained by requiring 
$\Delta \chi^2_{95\%}=3.84~(5.99)$ relative to the corresponding $\chi^2$ minimum for each case.

In what follows we explain the results presented in table \ref{table: ZpLimits} for 
the $Z^\prime$ models in table \ref{table: ZpCharges} in turn:

\begin{table}[t]
\begin{center}
{\footnotesize
\begin{tabular}{c c c c c c c c c c c c c c c}
\ctoprule
&\!\!\!\!\! &\multicolumn{13}{c}{\bf{\underline{$\bfmath{95\%}$ C.L. Electroweak Limits}}}\\
\!\!\!\bf{Model}\!\!\!&\!\!\!\!\! &\!$\Zpbf{ }$ \bf{alone}\!&\!\!\!\!\! &\multicolumn{2}{c}{$\Zpbf{ }\bfmath{+}\overline{\Zpbf{ }}$}&\!\!\!\!\! &\multicolumn{3}{c}{$\Zpbf{ }\bfmath{+\varphi+\Delta}$}&\!\!\!\!\! &\multicolumn{4}{c}{$\Zpbf{ }\bfmath{+}\overline{\Zpbf{ }}\bfmath{+\varphi+\Delta}$}\\
&\!\!\!\!\! &$\bfmath{M_{Z^\prime}}$&\!\!\!\!\! &$\bfmath{M_{Z^\prime}}$&$\bfmath{\frac{|g_{\overline{Z^\prime}}|}{M_{\overline{Z^\prime}}}}$&\!\!\!\!\! &$\bfmath{M_{Z^\prime}}$&$\bfmath{\frac{|\lambda_\varphi|}{M_{\varphi}}}$&$\bfmath{\frac{|\lambda_\Delta|}{M_{\Delta}}}$&\!\!\!\!\! &$\bfmath{M_{Z^\prime}}$&$\bfmath{\frac{|g_{\overline{Z^\prime}}|}{M_{\overline{Z^\prime}}}}$&$\bfmath{\frac{|\lambda_\varphi|}{M_{\varphi}}}$&$\bfmath{\frac{|\lambda_\Delta|}{M_{\Delta}}}$\\
 &\!\!\!\!\! &{\bf \scriptsize [GeV]}&\!\!\!\!\! &{\bf \scriptsize [GeV]}&\!{\bf \scriptsize [TeV$\bfmath{^{-1}}$]}\!&\!\!\!\!\! &{\bf \scriptsize [GeV]}&\multicolumn{2}{c}{{\bf \scriptsize [TeV$\bfmath{^{-1}}$]}}&\!\!\!\!\! &{\bf \scriptsize [GeV]}&\multicolumn{3}{c}{{\bf \scriptsize [TeV$\bfmath{^{-1}}$]}}\\
\cmidrule{1-1}\cmidrule{3-3}\cmidrule{5-6}\cmidrule{8-10}\cmidrule{12-15}
&\!\!\!\!\! &&\!\!\!\!\! &&&\!\!\!\!\! &&&&\!\!\!\!\! &&&&\\[-0.4cm]
\crowcolor$\Zpbf{\chi}$&\whitecell \!\!\!\!\! &1035~~(900)&\whitecell \!\!\!\!\! &856&$0.77$&\whitecell \!\!\!\!\! &869&$0.19$&$0.36$&\whitecell \!\!\!\!\! &475&1.2&0.35&\!\!\![0.07,0.46]\!\!\! \\
$\Zpbf{I}$&\!\!\!\!\! &1144~~(842)&\!\!\!\!\! &878&1.3&\!\!\!\!\! &993&-&0.34&\!\!\!\!\! &681&1.6&-&0.38\\
\crowcolor$\Zpbf{S}$&\whitecell \!\!\!\!\! &1162~~(871)&\whitecell \!\!\!\!\! &895&0.79&\whitecell \!\!\!\!\! &994&0.17&0.35&\whitecell \!\!\!\!\! &434&1.4&0.38&0.49\\
$\Zpbf{LR}$&\!\!\!\!\! &1206~~(959)&\!\!\!\!\! &605&1.9&\!\!\!\!\! &1234&0.14&0.33&\!\!\!\!\! &611&1.9&0.17&0.33\\
\crowcolor$\Zpbf{R}$&\whitecell \!\!\!\!\! &1146 (1006)&\whitecell \!\!\!\!\! &697&1.9&\whitecell \!\!\!\!\! &1146&$0.15$&-&\whitecell \!\!\!\!\! &451&$2.5$&$0.36$&-\\
\!\!\!$\Zpbf{B-L}$\!\!\!&\!\!\!\!\! &1261~~(971)&\!\!\!\!\! & & &\!\!\!\!\! &905&0.27&0.37&\!\!\!\!\! &&&&\\
&\!\!\!\!\! &&\!\!\!\!\! &&&\!\!\!\!\! &&&&\!\!\!\!\! &&&&\\[-0.375cm]
\cmrule
\!\!\!\bf{Model}\!\!\!&\!\!\!\!\! & &\!\!\!\!\! &\multicolumn{2}{c}{$\Zpbf{I}\bfmath{+}\Zpbf{\eta}$}&\!\!\!\!\! &\multicolumn{3}{c}{}&\!\!\!\!\! &\multicolumn{4}{c}{$\Zpbf{I}\bfmath{+}\Zpbf{\eta}\bfmath{+\varphi+\Delta}$}\\
&\!\!\!\!\! & &\!\!\!\!\! &$\bfmath{M_{Z^\prime_I}}$&$\bfmath{M_{Z^\prime_\eta}}$&\!\!\!\!\! & & & &\!\!\!\!\! &$\bfmath{M_{Z^\prime_I}}$&$\bfmath{M_{Z^\prime_\eta}}$&$\bfmath{\frac{|\lambda_\varphi|}{M_{\varphi}}}$&$\bfmath{\frac{|\lambda_\Delta|}{M_{\Delta}}}$\\
 &\!\!\!\!\! & &\!\!\!\!\! &{\bf \scriptsize [GeV]}&{\bf \scriptsize [GeV]}&\!\!\!\!\! & & & &\!\!\!\!\! &{\bf \scriptsize [GeV]}&{\bf \scriptsize [GeV]}&\multicolumn{2}{c}{{\bf \scriptsize [TeV$\bfmath{^{-1}}$]}}\\
\cmidrule{1-1}\cmidrule{5-6}\cmidrule{12-15}
&\!\!\!\!\! &&\!\!\!\!\! &&&\!\!\!\!\! &&&&\!\!\!\!\! &&&&\\[-0.4cm]
\crowcolor \!\!\!\!$\Zpbf{I},\Zpbf{\eta}$\!\!\!\!&\whitecell \!\!\!\!\! &\whitecell &\whitecell \!\!\!\!\! &1061&489&\whitecell \!\!\!\!\! &\whitecell&\whitecell&\whitecell&\whitecell \!\!\!\!\! &767&373&0.32&0.39\\
\cbottomrule
\end{tabular}
\caption{$95\%$ C.L. electroweak limits on the $Z^\prime$ masses for some of the most popular models. 
We compare the limits obtained in the single $Z^\prime$ scenario with those 
including extra $\overline{\Zp}$ vectors and scalars. See text for details and table \ref{table: ZpCharges} for the 
$\Zp$ and $\overline{\Zp}$ charges. For comparison, we also give in parentheses in the first column the most stringent limit from direct searches at the Tevatron and LHC \cite{Abe:1997fd,Erler:2010uy,Aad:2011xp}.}
\label{table: ZpLimits}}
\end{center}
\end{table}

\begin{itemize}
\item{$Z^\prime_\chi$:}{ The inclusion of the extra scalars suffices to pull the limit on $M_{Z^\prime}$ 
for the $\chi$ model below the direct LHC bound of $900\units{GeV}$, even in the absence of any other 
$Z^\prime$. In order to understand this, let us note that in this case the LEP 2 observables do not require 
an increase of the mass limit obtained from the $Z$-pole and low-energy data alone. On the contrary, they 
lower it (see table 5 in \cite{delAguila:2010mx}). This is because the effect of the $\chi$ model tends to 
increase the total $e^+ e^-\rightarrow \mt{had}$ cross section, which is $1.7~\sigma$ above the SM. Thus, 
the addition of new scalars allows one  to maintain this enhancement and compensate for the $Z^\prime_\chi$ 
contributions to leptonic processes, which are in good agreement with the SM predictions. Furthermore, 
they also help to reduce the $1.3~\sigma$ discrepancy with the electron weak charge extracted from M{\o}ller 
scattering. A similar limit is obtained if we consider the two $Z^\prime$ scenario obtained by introducing the 
$\overline{Z^\prime}_\chi$ in table \ref{table: ZpCharges}. Note that in this particular case we have chosen 
not to couple the $\overline{Z^\prime}_\chi$ to the RH leptons. Thus, we are not canceling any of the operators 
with RH quarks and leptons. This particular choice preserves large contributions to the hadronic cross section at LEP 2, 
while the cancellations prevent a significant discrepancy with the atomic parity violation data. Finally, 
when we include all the new particles, the limit on $M_{Z^\prime_\chi}$ is lowered to around 475 GeV. 

The effect of adding different particles is illustrated in figure \ref{Fig_Chi}. On the left we show the $95\%$ 
confidence region in the $M_{Z^\prime_\chi}$-$M_{\overline{Z^\prime}_\chi}/g_{\overline{Z^\prime}_\chi}$ plane 
from a fit to the model with two $Z^\prime$s alone (inner region), and with extra scalars in addition (outer region). 
On the right we show the corresponding regions in the $M_{Z^\prime_\chi}$-$M_{\varphi}/\lambda_{\varphi}$ 
plane (in this case the inner one corresponds to the fit to $Z^\prime_\chi$ alone plus the extra scalars). 
Apart from the scales, both figures look almost the same. Notice the significant correlation 
for low masses when we include all the particles at the same time. The correlation is less pronounced and the effect 
on the $M_{Z^\prime_\chi}$ bound smaller for each separate addition. 
In particular, there is no appreciable correlation between the $M_{Z^\prime_\chi}$ lower bound and $\varphi$, 
as observed in figure  \ref{Fig_Chi}, right panel. In this case its significant reduction is 
mostly due to the triplet scalar $\Delta$ contribution.
}
\begin{figure}[t]
\input{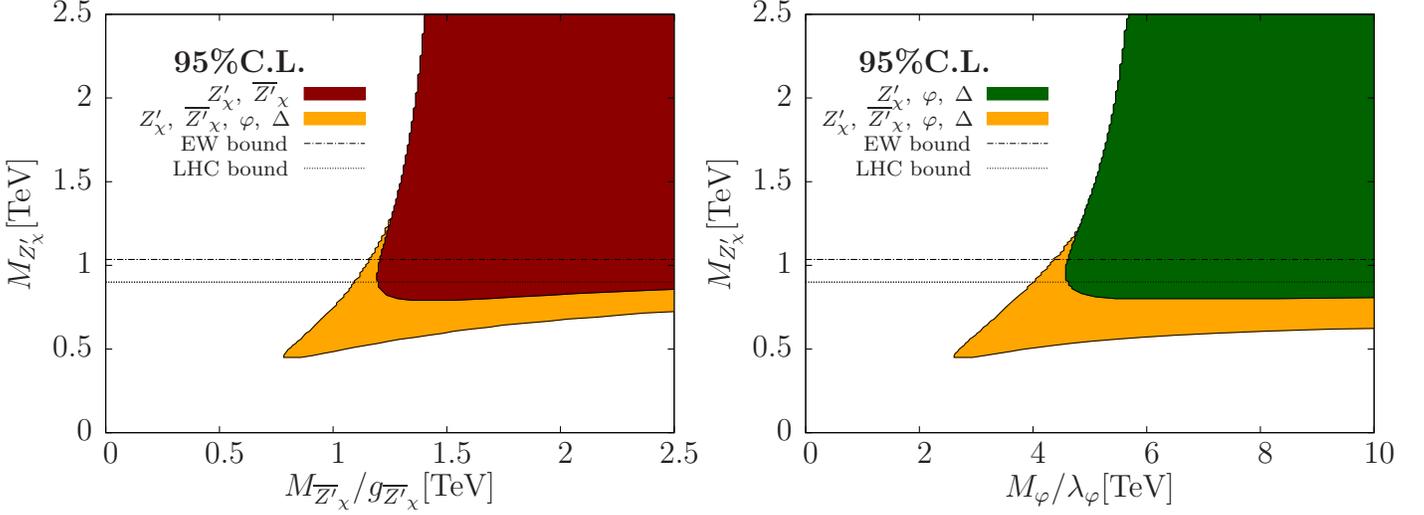}
\caption{(Left) $95\%$ C.L. confidence regions in the $M_{Z^\prime_\chi}$-
$(g_{\overline{Z^\prime}_{\chi}}/M_{\overline{Z^\prime}_{\chi}})^{-1}$ parameter space from a 
two-$Z^\prime$ fit  with and without including the scalars $\varphi$ and $\Delta$ (light [ocher] and dark [brown] 
solid regions, respectively). (Right) The same in the $M_{Z^\prime_\chi}$-
$(\lambda_{\varphi}/M_{\varphi})^{-1}$ plane from the fit to a $Z^\prime_\chi$ plus the scalars 
$\varphi$, $\Delta$ (dark [green] solid region), and the fit including the $\overline{Z^\prime}_\chi$ (light [ocher] solid region).}
\label{Fig_Chi}
\end{figure}
\item{$Z^\prime_I$:}{ As in the $Z_\chi^\prime$ case, the limit on the $Z^\prime_I$ mass can be  slightly 
reduced adding new scalars (only the scalar triplet in this case, since $Z^\prime_I$ does not couple to RH leptons). 
However, the new scalar is not enough to lower the EWPD limit below the LHC bound 
of $842 \units{GeV}$, because the electroweak bound is more stringent in this case. The $Z^\prime_I$ 
counterpart needs only to couple to LH leptons and RH $d$ quarks to attain a complete cancellation of all the four-fermion 
operators with no definite sign. The sole addition of the $\overline{Z^\prime}_I$ in table \ref{table: ZpCharges} lowers the 
limit around 100 GeV below that obtained with the scalar triplet. However,  a complete cancellation of all 
four-fermion contributions is not possible when both $\overline{Z^\prime}_I$ and $\Delta$ are included. This is  because, 
even if we choose completely general couplings for $\Delta$, $\mu$ decay constraints on the electron-muon couplings 
prevent the cancellation of the $Z^\prime$ contribution to the operator with two LH electrons and two LH muons. 
Still, the combined scenario in the right columns of table \ref{table: ZpLimits} suffices to lower the electroweak limit 
significantly below the direct searches bound.

We can also lower the $M_{Z_I^\prime}$ limits with 
a second $Z^\prime$ within $E_6$, $Z^\prime_\eta$. 
(See table \ref{table: ZpLimits}.) 
Actually, the charges of the $\eta$ and $I$ models are orthogonal 
within this group. When we combine these two $Z^\prime$s with the (two) extra scalars, we can lower the limit on 
$M_{Z^\prime_I}$ below the LHC bound of $842 \units{GeV}$. 
This can be seen in figure \ref{Fig_EtaI}. However, 
this limit occurs in correlation with a low $Z^\prime_\eta$ mass and this is excluded below $910\units{GeV}$ by Tevatron searches.\footnote{We assume these limits still apply in the two-$Z^\prime$ models, which is a good approximation if the 
resonances are narrow enough to be distinctively separated.} Taking this into account, the limit on $M_{Z^\prime_I}$ is still slightly below the LHC bound. 
}
\begin{figure}[t]
\input{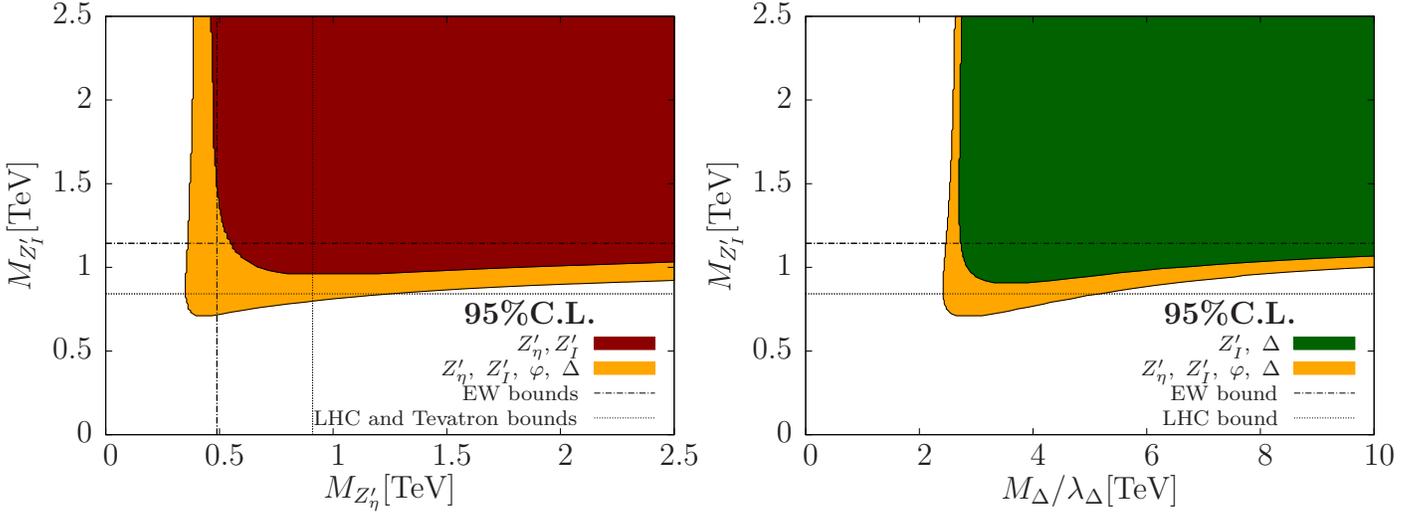}
\caption{(Left) $95\%$ C.L. confidence regions in the $M_{Z^\prime_I}$-$M_{Z^\prime_\eta}$ parameter space 
from the two-$Z^\prime$ fit  with and without including the scalars $\varphi$ and $\Delta$ (light [ocher] and dark [brown] solid 
regions, respectively). (Right) The same in the $M_{Z^\prime_I}$-$(\lambda_{\Delta}/M_{\Delta})^{-1}$ plane 
from the fit to the $Z^\prime_I$ plus the scalar $\Delta$ (dark [green] solid region) and from the fit also including 
$Z^\prime_\eta$ and $\varphi$ (light [ocher] solid region).}
\label{Fig_EtaI}
\end{figure}
\item{$Z^\prime_S$:}{ The $Z^\prime_S$ charges have a pattern rather similar to those of the $Z^\prime_\chi$. 
Hence, we choose similar charges for its counterpart $\overline{Z^\prime}_S$. Then,  a similar 
discussion regarding the scalar additions and the combined scenario including all extra particles also applies, as can be seen in table \ref{table: ZpLimits}.
}
\item{$Z^\prime_{LR}$:}{ The limit on the $Z^\prime_{LR}$ mass cannot be relaxed by introducing 
extra scalars. The reason is that for this model the LEP 2 constraints are dominated by the $e^+ e^-\rightarrow \mt{had}$ data 
and the effect of the $Z^\prime$ is to reduce the total cross section relative to the SM, increasing the discrepancy 
with experiment.\footnote{Although it may seem surprising that in this case the $95\%$ C.L. on the
 $Z^\prime_{LR}$ mass in
table~\ref{table: ZpLimits} is slightly higher when adding new scalars (parameters), this is so
because this limit is relative to the corresponding new minimum. This is deeper due to the scalar 
contributions which do not decouple near this point, then redefining the probability distribution.}
We can ameliorate 
these restrictions, as well as those from low-energy data (in particular those from atomic parity violation experiments), 
by introducing a $\overline{Z^\prime}_{LR}$ with charges designed to cancel the left-right model contributions to operators with 
two leptons and two quarks. This addition alone suffices to lower the limit to half the electroweak bound in the single 
$Z^\prime$ case. We find no improvement in this case when we also add extra scalars.
}
\item{$Z^\prime_R$:}{ Similar to the LR model, the limits on $Z^\prime_R$ can be drastically 
reduced adding a $\overline{Z^\prime}_R$. Also as in the LR case, the cancellation of the purely leptonic contributions by 
adding extra scalars alone leaves the limits intact. However, a significant improvement is possible when we combine 
the two additions. Since the
$Z^\prime_R$
only couples to RH fermions a complete cancellation of all four-fermion
contributions would be possible with the addition of the second
$\overline{Z^\prime}_R$ and of scalar singlets $\varphi_\ell$ with couplings
properly chosen. For our specific choice of
scalar couplings, however, this cancellation is incomplete. Hence, we can
find a
$95\%$ C.L. limit on $M_{Z^\prime_R}$. As can be observed by comparing
Eqs. (\ref{B4lops}) and (\ref{singlet op}), a perfect cancellation of the
leptonic four-fermion
operators requires that the scalar couplings satisfy the equality
\be
\lambda_{\varphi_e}^{ee}/\sqrt{2}=\lambda_{\varphi_\mu}^{e\mu}=\lambda_{\varphi_\mu}^{\mu
e}=\lambda_{\varphi_\tau}^{e\tau}=\lambda_{\varphi_\tau}^{\tau e}.
\label{scalchoice2}
\ee
In such a case
there is a flat direction in the parameter space, allowing for arbitrary
$M_{Z^\prime_R}$ values by adjusting the other extra parameters. 
This is illustrated in figure \ref{Fig_R}, which is analogous to 
figure \ref{Fig_Chi} for $Z^\prime_\chi$, but with the
scalar coupling choice in (\ref{scalchoice2}). We must emphasize, however,
that in the effective Lagrangian approach used here the fit  only makes sense
for $M_{Z^\prime_R}$ above the maximum LEP 2 energies $\sim 209 \units{GeV}$.
}
\begin{figure}[t]
\input{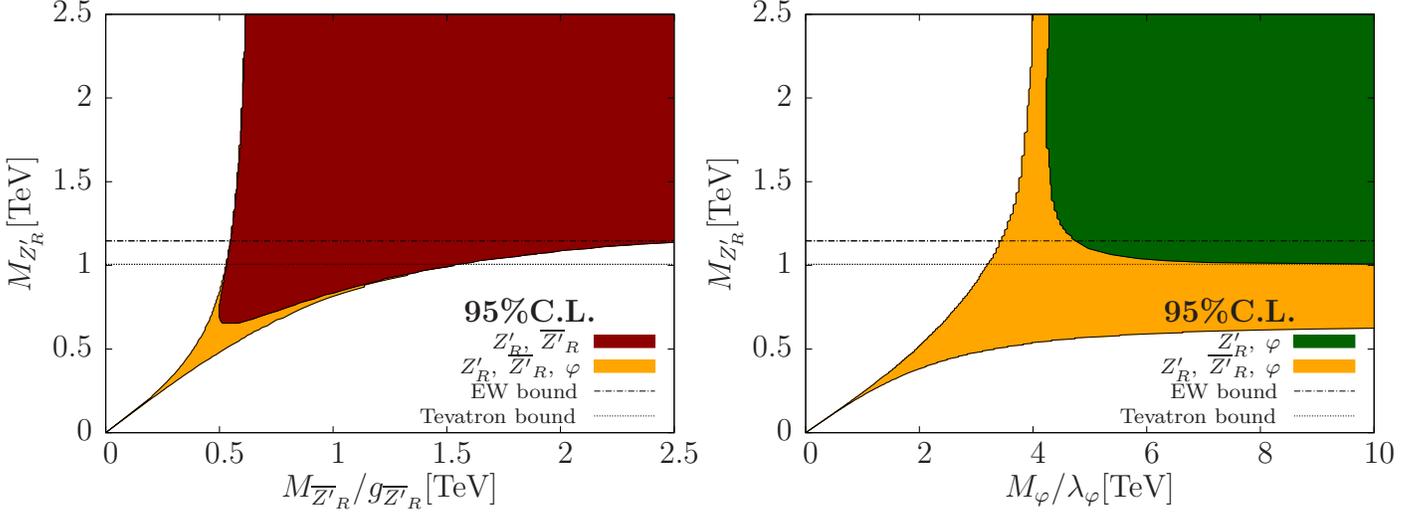}
\caption{(Left) $95\%$ C.L. confidence regions in the $M_{Z^\prime_R}$-
$(g_{\overline{Z^\prime}_{R}}/M_{\overline{Z^\prime}_{R}})^{-1}$ parameter space from a 
two-$Z^\prime$ fit  with and without including the scalars $\varphi$ (light [ocher] and dark [brown] 
solid regions, respectively). In this case, the scalar couplings are specifically chosen to attain a perfect cancellation
of the purely leptonic $Z^\prime_R$ effects. See text for details. (Right) The same in the $M_{Z^\prime_R}$-
$(\lambda_{\varphi}/M_{\varphi})^{-1}$ plane from the fit to a $Z^\prime_R$ plus the scalars 
$\varphi$ (dark [green] solid region), and the fit including the $\overline{Z^\prime}_R$ (light [ocher] solid region).}
\label{Fig_R}
\end{figure}
\item{$Z^\prime_{B-L}$:}{ The limit on the $Z^\prime_{B-L}$ mass is to a large extent determined by purely 
leptonic LEP 2 data. Thus, we do not find any $\overline{Z^\prime}$ that can lower this limit. On the other hand, the addition 
of new scalars does allow for a $M_{Z^\prime_{B-L}}$ limit around $350 \units{GeV}$ lower than in the single $Z^\prime_{B-L}$ 
case.
}
\end{itemize} 

The corresponding contours for the $Z^\prime_S$, $Z^\prime_{LR}$, $Z^\prime_R$ (for our standard choice of $\varphi$ couplings), and $Z^\prime_{B-L}$
 are analogous to figures \ref{Fig_Chi} and \ref{Fig_EtaI} for the $Z^\prime_\chi$ and $Z^\prime_I$.

\begin{figure}[t]
\input{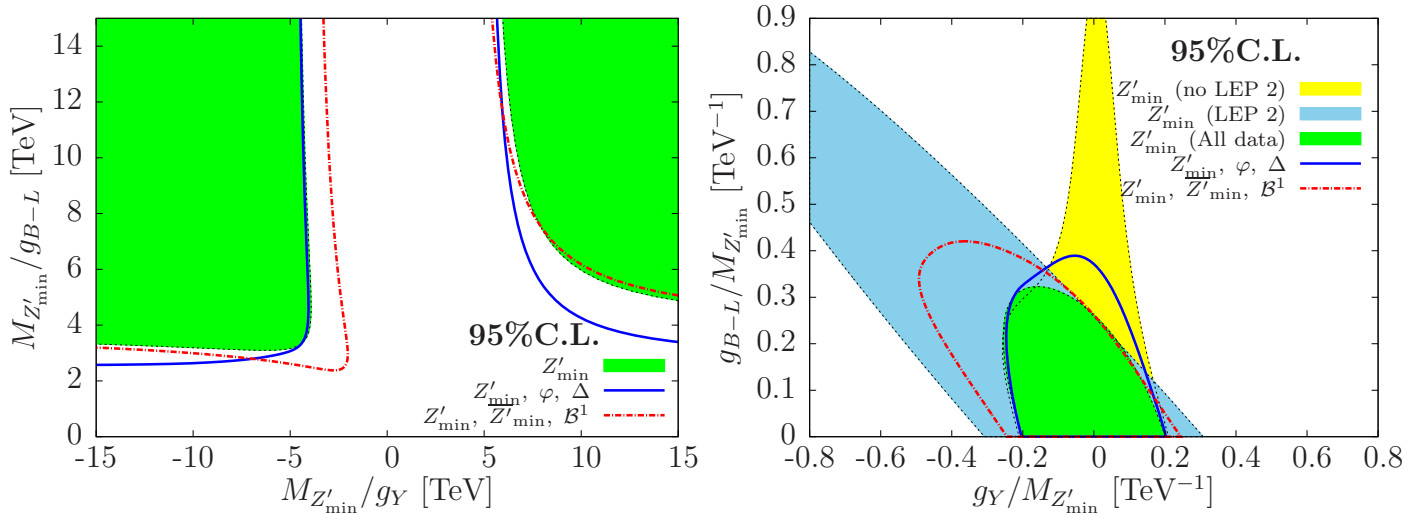}
\caption{(Left) $95\%$ C.L. confidence regions in the $M_{Z^\prime_\mt{min}}/g_Y$-$M_{Z^\prime_\mt{min}}/g_{B-L}$ parameter 
space for the minimal $Z^\prime$ models (dark [green] solid region) and their extensions including also the scalars $\varphi$ and $\Delta$ 
(solid [blue] line) or a ${\cal B}^1$ together with a $\overline{Z^\prime}$ with {\em mirror} charges (dot-dashed [red] line). 
(Right) The same in the $g_Y/M_{Z^\prime_\mt{min}}$-$g_{B-L}/M_{Z^\prime_\mt{min}}$ plane. We also include the regions 
corresponding to the fit to EWPD without LEP 2 $e^+e^-\rightarrow \bar{f}f$ data (vertical [yellow] band) and to LEP 2 data only (diagonal 
[blue] band) for the $Z^\prime_\mt{min}$ model alone.}
\label{Fig_Minimal}
\end{figure}

Finally, we discuss the {\em minimal \zp models} studied in \cite{Salvioni:2009mt}.
Their charges are a linear combination of the hypercharge $Y$ and $B-L$. Thus, this case is fully characterized by the $Z^\prime$ 
mass $M_{Z^\prime_\mt{min}}$ and the two coupling constants $g_Y$ and $g_{B-L}$ defining its current. Following 
\cite{Salvioni:2009mt} we normalize these constants in such a way that the fermionic current coupling to the 
$Z^\prime_\mt{min}$ is given (before mixing with the $Z$) by 
\be
J_{Z^\prime_\mt{min}}^\mu\supset \sqrt{g^2+g^{\prime~2}}\sum_\psi \left[g_Y Y_\psi+g_{B-L} (B-L)_\psi\right]\overline{\psi}\gamma^\mu \psi.
\label{Jmin}
\ee
Thus, the fit only constrains the ratios $g_Y/M_{Z^\prime_\mt{min}}$ and $g_{B-L}/M_{Z^\prime_\mt{min}}$. 
In figure \ref{Fig_Minimal} we depict the $95\%$ confidence regions using two different parameterizations. 
On the left we draw the $M_{Z^\prime_\mt{min}}/g_Y$ - $M_{Z^\prime_\mt{min}}/g_{B-L}$ plane to facilitate the 
comparison with previous cases, and on the right the 
$g_Y/M_{Z^\prime_\mt{min}}$ - $g_{B-L}/M_{Z^\prime_\mt{min}}$ plane as done in  
\cite{Salvioni:2009mt}. In this class of models 
the relative sign between $g_Y$ and $g_{B-L}$ is physical.
The limits are in general more stringent than the ones for the popular models above
 because by construction the $Z^\prime$ mixing with the $Z$ boson is 
not a free parameter and is nonvanishing. 
Taking into account the different normalization used, which stands for a multiplicative 
factor $\sqrt{5/3} {g^\prime} / \sqrt {g^2 + {g^\prime}^2} \approx 0.6$, the lower limits 
in figure \ref{Fig_Minimal}, left panel, are almost a factor $\sim 3$ larger than those in the  previous figures.

The results can be better visualized in figure \ref{Fig_Minimal}, right panel, where the limits  
are plotted as a function of $g_Y/M_{Z^\prime_\mt{min}}$ and $g_{B-L}/M_{Z^\prime_\mt{min}}$. 
As in \cite{Salvioni:2009mt}, we also draw the constraints from different data sets. 
In this class of models the addition of new scalars has a limited effect.  
The corresponding $95\%$ confidence region, which is delimited by the solid (blue) contours in figure \ref{Fig_Minimal}, is not much larger than that of the minimal model alone. At any rate, due to the cancellation of the purely leptonic effects, the LEP 2 bounds can be somewhat relaxed and the allowed region is enlarged along the band determined by the $Z$-pole, $M_W$ and low-energy data (among others). When we include
 the new gauge bosons ${\cal B}^1$ and $\overline{Z^\prime}_\mt{min}$,\footnote{${\cal B}^1$ is a fermiophobic singlet vector boson with hypercharge $Y=1$, 
following the notation introduced in \cite{delAguila:2010mx}; whereas $\overline{Z^\prime}_\mt{min}$ is 
a $Z^\prime$ with {\em mirror} minimal couplings, as described in section \ref{SVFop}. These additions allow for a complete cancellation of the $Z$-$Z^\prime_\mt{min}$ mixing effects, 
as discussed in sections \ref{Obliqueop} and \ref{SVFop}. Of course, it is also possible to avoid these
effects by allowing more general Higgs structures, as also described in those sections.}
the constraints from the $Z$-pole data and the observables sensitive to oblique effects can be significantly relaxed. Thus, the extended confidence region, delimited by the dot-dashed (red) contours in the figure, opens along the band allowed by LEP 2 data. The low-energy constraints, however, prevent one from obtaining a much larger region. 
On the other hand,  (even) more contrived constructions including additional $Z^\prime$s, 
in order to relax the remaining low-energy and LEP 2 hadronic constraints, may allow for larger regions.


\section{Conclusions}
\label{Conclusions}

New particles can manifest themselves as resonances at large colliders, or 
indirectly as deviations from the SM predictions in some 
observables. 
In this sense the Tevatron and LHC are complementary tools 
to EWPD for new physics searches, although none of them has 
provided significant evidence for physics beyond the SM yet.  
It is well known, however, that the quite good agreement of 
the SM predictions with EWPD implies that {\em simple} new physics 
is banished above the TeV scale, near the LHC reach. It is then 
important to investigate if more {\em complex} scenarios 
allow for small contributions to EWPD but relatively light 
new particles. This question is especially relevant to guide 
LHC searches. 

In this paper we have addressed this question for extra neutral gauge bosons. 
We have discussed in turn several popular $Z^\prime$ models 
based on $E_6$ and with EWPD mass limits above present 
direct bounds from the Tevatron and LHC. 
In particular, we have studied which additions of extra vector bosons 
and scalars can cancel their main contributions to EWPD.
Using the effective Lagrangian approach, which is 
especially suited for comparing or combining different extensions of the SM, 
one can decide  if there is a choice 
of couplings which may partially cancel the large contributions 
of any given extra gauge boson. 
We found that in all cases the EWPD bounds on the $Z^\prime$ 
masses can be lowered below the present direct limits, 
although with specific additions designed for this purpose.\footnote{We also require that 
the new particles have not been observed. In particular, that the custom $\overline{Z^\prime}$
has a dilepton production cross section at large hadron colliders smaller than that of the $Z^\prime$.} 
Otherwise, the EWPD limits appear to be robust.

We emphasize that the interest of the analysis presented here
goes beyond the popular $Z^\prime$ examples considered in section \ref{section: num_results}.
Indeed, the methods and results in section \ref{section: Evading EWL} are valid for arbitrary $Z^\prime$
bosons with universal couplings, and the generalization to nonuniversal and flavour-changing couplings
is straightforward.

If dilepton resonances are not found when more LHC data are available, 
the direct limits will eventually overcome those derived from EWPD for definite 
$Z^\prime$s coupling to quarks and leptons, as long as the couplings are of the electroweak
or GUT order. However, for large couplings the EWPD require large $Z^\prime$
masses, which may be beyond the reach of the LHC (at least at 7 TeV). Indeed, $Z^\prime$ production
is suppressed at hadron colliders for large masses, due to the energy dependence of the parton distribution functions. This suppression is stronger than the $1/M_{Z^\prime}$ scaling of EWPD limits. Thus, for large couplings the EWPD bounds may remain competitive.

In the opposite limit, leptophobic $Z^\prime$ bosons \cite{del Aguila:1986ez} can be very light,
since they evade Tevatron and LHC Drell-Yan bounds, and are not constrained by EWPD if they do 
not mix with the $Z$ boson. These $Z^\prime$ bosons have been recently 
invoked \cite{Jung:2009jz,Buckley:2011vc} to account for
Tevatron anomalies in the top forward-backward asymmetry \cite{Aaltonen:2011kc} (other
direct constraints apply in this case \cite{AguilarSaavedra:2011vw}) and $W+jj$ 
distribution \cite{Aaltonen:2011mk}. Here
we have shown that the EWPD constraints can also be evaded in models of leptophobic $Z^\prime$ bosons
with nonvanishing mixing. We note in passing that, because only the $W$ mass and $Z$-pole observables
are relevant in this case, the effective Lagrangian approach can be accurate enough for a leptophobic $Z^\prime$ with a mass $\sim 150\units{GeV}$.

\section*{Acknowledgements}
It is a pleasure to thank Antonio Delgado for reading and discussing the manuscript.
This work has been partially supported by 
MICINN (FPA2006-05294 and FPA2010-17915) 
and by Junta de Andaluc\'{\i}a (FQM 101, FQM 3048 and FQM 6552). 
The work of J.B. has been supported in part by the U.S. National Science Foundation under Grant PHY-0905283-ARRA, and that of P.L. by an IBM Einstein Fellowship and by NSF grant PHY--0969448.



\end{document}